\documentclass[10pt]{article}
\usepackage{amssymb,euscript,bbold}
\usepackage{amsfonts}
\usepackage{amssymb}
\usepackage{mathrsfs}
\usepackage{graphicx}
\usepackage{slashed}
\usepackage{amsthm}
\usepackage{amsmath}
\newtheorem{theorem}{Theorem}

\newtheorem{proposition}{Proposition}

\newtheorem*{definition}{Definition}

\UseRawInputEncoding
\newcommand{\be}{\begin{equation}}
\newcommand{\ee}{\end{equation}}
\newcommand{\ba}{\begin{eqnarray}}
\newcommand{\ea}{\end{eqnarray}}

\newcommand{\op}[1]{\operatorname{#1}} 
\usepackage{xcolor}
\usepackage{bbold} 
\usepackage{hyperref}
\usepackage{tabularx}

\begin{document}
\input{epsf}

\begin{flushright}
KCL-PH-TH/2023-72
\end{flushright}
\begin{flushright}
\end{flushright}
\begin{center}
\Large{Ricci Flat Metrics, Flat Connections and $G_2$-manifolds.}\\
\bigskip
\large{B.S. Acharya} and 
\large{D.A. Baldwin}\\
\smallskip\normalsize{\it
Abdus Salam International Centre for Theoretical Physics, Strada Costiera 11, 34151, Trieste, Italy}\\

and

{\it Department of Physics, Kings College London, London, WC2R 2LS, UK}\\
\end{center}

\bigskip
\begin{center}
{\bf {\sc Abstract}}

Inspired by considerations in $M$-theory, we prove the equivalence between the moduli spaces of (suitably complexified) torsion free $G_2$-structures on 7-manifolds which are families of hyperK\"ahler ALE 4-manifolds fibered over compact flat 3-manifolds $T^3/K$ and certain moduli spaces of flat $G^{\mathbb{C}}_{ADE}$ connections on $T^3/K$. 
\end{center}

\newpage

\section*{Introduction}

Ricci flat Riemannian manifolds with special holonomy have played an important role in superstring/$M$-theory as models of extra dimensions of space. In particular, special kinds of singularities which occur in the moduli spaces of such spaces have played an important role, as they often have a physical description in terms of a quantum field theory in lower dimensions, typically a gauge theory. This then leads to perhaps surprising connections between moduli spaces of Einstein metrics and moduli spaces in lower dimensional gauge theory. This note will describe examples of such connections within the specific context of certain manifolds with holonomy group contained in $G_2$.

This work was inspired by the study of $M$-theory compactifications on so-called Joyce-Karigiannis manifolds \cite{joyce2021new} which are constructed from certain $G_{2}$-holonomy spaces with orbifold singularities that admit topologically distinct desingularisations and Ricci flat $G_2$-holonomy metrics. 
The examples of interest here are Ricci flat metrics on 7-manifolds which are fibered by hyperK\"ahler 4-manifolds over compact flat 3-manifolds. In \cite{Acharya:2023xlx}, a correspondence between the moduli space of $G_{2}$ holonomy metrics on these 7-manifolds, the moduli space of complex flat connections on the 3-manifold and the vacuum moduli space of the four dimensional theory arising from the $M$-theory compactification was discovered and used to physically interpret the topologically distinct desingularisations as Coulomb and Higgs branches of the four dimensional gauge theory. Other notable aspects were considered by Barbosa \cite{barbosa2019harmonic,barbosa2019thesis} and further related works are \cite{pantev2011hitchin,Barbosa:2019bgh,Hubner:2020yde}. 

The purpose of this note is to make the correspondences in \cite{Acharya:2023xlx} more precise. We
will derive an isomorphism between the moduli space of complex flat $ADE$ connections on all orientable compact flat 3-manifolds $T^{3}/K$ and a suitably complexified moduli space of Ricci-flat metrics on certain complete 7-manifolds with $G_2$-structures, $(T^{3}\times X_{ADE})/K$, with $X_{ADE}$ a 4-manifold admitting a $K$-invariant hyperK\"ahler metric. The theorems we prove, in a sense, give a complete description of the classical moduli spaces of these physical systems or, in more physical terms, a complete description of the semi-classical vacuum structure.

\section*{Flat Connections and Ricci flat Metrics}

\subsection*{\it Flat Connections on Flat Manifolds}

We begin by a brief description of the moduli space of flat $G_{ADE}^{\mathbb{C}}$ connections\footnote{Throughout this paper, the maximal compact subgroup of $G_{ADE}^{\mathbb{C}}$ is $G_{ADE} = SU(N), Spin(2N), E_{6,7,8}.$} modulo gauge transformations on the 3-torus, $T^{3} \cong \mathbb{R}^3/\Lambda$, denoted $\mathcal{M}(T^{3},G_{ADE}^{\mathbb{C}})$  These are given by homomorphisms ($\mathsf{R}$) from $\pi_{1}(T^{3}) = \Lambda \cong \mathbb{Z}^3$ to $G_{ADE}^{\mathbb{C}}$ modulo conjugation by $G_{ADE}^{\mathbb{C}}$.
Up to conjugation, this is thus a commuting triple $\mathsf{R}(g_{i})$ of elements of $G_{ADE}^{\mathbb{C}}$ in the representation $\mathsf{R}$.
In general, $\mathcal{M}(T^{3},G_{ADE}^{\mathbb{C}})$ will have multiple components \cite{kac-smilga:in2p3-00195525,borel-friedman-morgan1999commuting,Keurentjes:2000mk,Witten_indexof4dgaugethy:2000nv}. Here it will suffice to describe the component containing the trivial connection, $R(g_i) = \mathbb{1}$, denoted by $\mathcal{M}_{0}(T^{3},G_{ADE}^{\mathbb{C}})$. In this case all three elements belong to the same maximal torus $\mathcal{T}^{\mathbb{C}} \equiv \mathbb{C}^{*}\otimes \mathfrak{h}_{ADE} \subset G_{ADE}^{\mathbb{C}}$, with $\mathfrak{h}_{ADE}$ the Cartan subalgebra of $G_{ADE}$. Since conjugation into the $\mathcal{T}$ does not fix the non-identity connected gauge transformations preserving  $\mathcal{T}$, one is left with residual gauge transformations by elements of the Weyl group, $\mathcal{W}_{ADE}$ of $G_{ADE}$, hence we have that

\begin{equation}
\mathcal{M}_{0}(T^{3},G_{ADE}^{\mathbb{C}}) \equiv (\mathcal{T^{\mathbb{C}}})^3/\mathcal{W}_{ADE}
\end{equation}

Next we consider flat connections on an orientable compact flat 3-manifold $M^3_K = T^{3}/K$. Since $M_{K}^{3}$ is smooth and flat, $K$ is a finite group of isometries acting freely on $T^3$.
Such manifolds are classified and the possibilities are $K=\mathbb{1}, \mathbb{Z}_{2},\mathbb{Z}_{3},\mathbb{Z}_{4},\mathbb{Z}_{6},\mathbb{Z}_{2}\times \mathbb{Z}_{2}$. Their fundamental groups are described explicitly in table \ref{tab:relations table}.

The fundamental group $\pi_1(M^3_K)$ fits into an exact sequence of groups:
\begin{equation}
0  \rightarrow \Lambda \rightarrow \pi_1 (M^3_K) \rightarrow K \rightarrow 1 
\end{equation}
A flat connection on $M^3_K$ is, up to conjugation, a homomorphism $\op{Hom}(\pi_1 (M^3_K), G_{ADE}^{\mathbb{C}})$ and the above sequence implies the following,
\begin{equation}
    \op{Hom}(K,G_{ADE}^{\mathbb{C}})\to \op{Hom}(\pi_{1}(M_{k}^{3}),G_{ADE}^{\mathbb{C}})\to \op{Hom}(\Lambda,G_{ADE}^{\mathbb{C}})
\end{equation}

One can thus view a flat connection on $M^3_K$ as a flat connection on $T^3$ which is $K$-invariant, as the elements of $\op{Hom}(K,G_{ADE}^{\mathbb{C}})$ act on $\op{Hom}(\Lambda,G_{ADE}^{\mathbb{C}})$.
Thus, the moduli space of flat $G_{ADE}^{\mathbb{C}}$ connections on $T^{3}/K$ is given by the $K$-invariant subspace of the flat connections on $T^{3}$,

\begin{equation}
    \mathcal{M}(T^{3}/K,G_{ADE}^{\mathbb{C}}) = \left(\mathcal{M}(T^{3},G_{ADE}^{\mathbb{C}})\right)^{K}
\end{equation}

We will often be interested in the $K$-invariant flat connections which come from the identity connected flat connections on $T^3$:
\begin{equation}
    \left(\mathcal{M}_{0}(T^{3},G_{ADE}^{\mathbb{C}})\right)^{K}
\end{equation}

In table \ref{tab:relations table} we give a description of the fundamental group relations for all $K$ which must therefore be satisfied by any flat connection $\mathsf{R} \in \op{Hom}(\pi_1 (M^3_K), G_{ADE}^{\mathbb{C}})$. From here we see that, if $\mathsf{R}(g_{1,2,3})$ all belong to the same maximal torus $\mathcal{T}$,
\begin{equation}
    \mathsf{R}(K): \mathcal{T} \rightarrow \mathcal{T}
\end{equation}
Since $\mathsf{R}(K)$ preserves $\mathcal{T}$ we can regard it as representing  $\mathsf{N}(\mathcal{T})$, the normaliser of this maximal torus.
However, the action of $\mathcal{T}$ on $\mathsf{R}(K)$ by conjugation is a gauge transformation which reduces the independent degrees of freedom to those of the Weyl group: $\mathsf{N}(\mathcal{T})/\mathcal{T} = \mathcal{W}_{ADE}$. 
\begin{theorem}
Suppose $\mathsf{R} \in \op{Hom}(\pi_1 (M^3_K), G_{ADE}^{\mathbb{C}})$ is a flat ADE-connection whose restriction to $\mathsf{R}(\Lambda) \subset \mathcal{M}_{0}(T^{3},G_{ADE}^{\mathbb{C}})$. Then all such flat connections are in one-to-one correspondence with $\op{Hom}(K, \mathcal{W}_{ADE}$). 
\end{theorem}
{\it Proof.} Given a homomorphism  $\mathsf{R}'(K)$ in $ \op{Hom}(K, \mathcal{W}_{ADE})$, one has a $K$-action on $\mathcal{M}_{0}(T^{3},G_{ADE}^{\mathbb{C}})$ and the $K$-invariant subspace, taken together with $\mathsf{R}'(K)$ is a flat connection on $M^3_K$. Conversely, given a flat connection on $M^3_K$ with $\mathsf{R}(\Lambda) \subset \mathcal{M}_{0}(T^{3},G_{ADE}^{\mathbb{C}})$, $\mathsf{R}(K) \subset \mathcal{N}(T)$ is gauge equivalent to a subgroup of $\mathcal{W}_{ADE}$. $\blacksquare$

One can also treat all flat connections on $M^3_K$, not just those that restrict to identity connected connections on $T^3$. Given a non-identity flat $ADE$-connection $A$ on $T^3$, one can consider the centraliser (or commutant) $C(A)$. This is the subgroup of $G_{ADE}$ commuting with $A$. Then, what we would like is for Theorem 1 to apply, but replacing $\mathcal{W}_{ADE}$ with the Weyl group of $C(A)$. This is in fact true, but only if $A$ is $K$-invariant.

\begin{table}
    \centering
    \begin{tabular}{c|c}
        $K$ & Relations \\
        \hline
        $\mathbb{1}$ & $g_{i}g_{j}=g_{j}g_{i},\qquad i,j\in \{1,2,3\}$ \\
        \hline
        $\mathbb{Z}_{2}$ & $g_{\beta}\,g_{1,2}\,g_{\beta}^{-1}=g_{1,2}^{-1},\qquad g_{\beta}\,g_{3}\,g_{\beta}^{-1}=g_{3},\qquad g_{\beta}^{2}=g_{3}$\\
        \hline
        $\mathbb{Z}_{3}$ & $g_{\beta}\,g_{1}vg_{\beta}^{-1}=g_{2},\qquad g_{\beta}\,g_{2}\,g_{\beta}^{-1}=g_{1}^{-1}\,g_{2}^{-1}$,\\
        & $g_{\beta}\,g_{3}\,g_{\beta}^{-1}=g_{3},\qquad g_{\beta}^{3}=g_{3}$\\
        \hline
        $\mathbb{Z}_{4}$ & $g_{\beta}\,g_{1}\,g_{\beta}^{-1}=g_{2}^{-1},\qquad g_{\beta}\,g_{2}\,g_{\beta}^{-1}=g_{1}$,\\
        & $g_{\beta}\,g_{3}\,g_{\beta}^{-1}=g_{3},\qquad g_{\beta}^{4}=g_{3}$\\
        \hline
        $\mathbb{Z}_{6}$ &$g_{\beta}\,g_{1}\,g_{\beta}^{-1}=g_{1}\,g_{2},\qquad g_{\beta}\,g_{2}\,g_{\beta}^{-1}=g_{1}^{-1}$,\\
        & $g_{\beta}\,g_{3}\,g_{\beta}^{-1}=g_{3},\qquad g_{\beta}^{6}=g_{3}$\\
        \hline
        $\mathbb{Z}_{2}\times \mathbb{Z}_{2}$ & $g_{\beta_{i}}\,g_{i}\,g_{\beta_{i}}=g_{i},\quad g_{\beta_{i}}^{2} = g_{i},$\\
        & $g_{\beta_{i}}\,g_{j}\,g_{\beta_{i}}^{-1}=g_{j}^{-1},\quad i\neq j,\quad i,j\in\{1,2,3\}$\\
        &$g_{\beta_{3}}=g_{1}\,g_{\beta_{2}}\,g_{\beta_{1}}$
    \end{tabular}
    \caption{The fundamental group relations for $T^{3}/K$ for all $K$. In all cases the elements $g_{1,2,3}$ mutually commute.}
    \label{tab:relations table}
\end{table}

We arrive at the following conclusion: the number of branches of the moduli space of flat $G_{ADE}^{\mathbb{C}}$ connections on $T^3/K$ satisfying the conditions of Theorem 1 is the same as the number of inequivalent homomorphisms from $K$ to $\mathcal{W}_{ADE}$.

\subsection*{\it Ricci flat ALE metrics and $G_2$-structures.}

Having classified flat $ADE$ connections on $M^3_K$ we now turn to proving the equivalence to certain moduli spaces of Einstein 7-manifolds with torsion free $G_2$-structures as predicted by $M$-theory. The idea is that $M$-theory on 7-manifolds of the form $X^7 = (T^{3}\times X_{ADE})/K$ where $X_{ADE} = \widetilde{\mathbb{C}^{2}/\Gamma_{ ADE}}$ is a smooth resolution of the orbifold admitting a Ricci flat, hyperK\"ahler, asymptotically locally Euclidean (ALE) metric, has a moduli space which is identical to the moduli space of flat $ADE$ connections on $M^3_K$ whose restriction to $T^3$ lies in $\mathcal{M}_{0}(T^{3},G_{ADE}^{\mathbb{C}})$.

In more detail, we consider

\begin{equation}\label{eq:T3 times Ak mod Z2}
   X= \frac{(T^{3}\times X_{ADE})}{K}
\end{equation}
with a Ricci flat metric 
\begin{equation}
    g = d\vec{y}^{2} + h
\end{equation} 
and a $K$-invariant torsion free $G_2$-structure
\begin{equation}
    \varphi = dy^{1}dy^{2}dy^{3} + d\vec{y}\cdot\vec{\omega},
\end{equation}
Here, $\vec{y}$ are Euclidean coordinates on $T^3$, $h$ is a $K$-invariant hyperK\"ahler metric on $X_{ADE}$ and $\vec{\omega}$ define the hyperK\"ahler structure on $X_{ADE}$.
Any such metric corresponds to a $K$-invariant desingularisation of the flat 7-orbifold
$X_0= \frac{(T^{3}\times \mathbb{C}^{2}/\Gamma_{ADE})}{K}$. In general, as we discussed in \cite{Acharya:2023xlx}, there will be multiple topologically distinct desingularisations of the same singularity, leading to multiple components to the moduli space of Ricci flat metrics.

We begin by desribing the moduli space of Ricci flat metrics for $K=\mathbb{1}$.

\begin{proposition}[Kronheimer \cite{Kronheimer}]
The moduli space of ALE hyperK\"ahler metrics on $X_{ADE}$ is $\mathcal{M}_{ALE} = (\mathbb{R}^3 \otimes \mathfrak{h})_{ADE}/\mathcal{W}_{ADE}$. 
\end{proposition}
From this it follows that:
\begin{proposition}
The moduli space of $SU(2)$ holonomy metrics on $T^3 \times X_{ADE}$, for a fixed flat metric on $T^3$ is $H_c^3(T^3 \times X_{ADE}, \mathbb{R})/\mathcal{W}_{ADE}$. 
\end{proposition}
This follows as the space of $SU(2)$-holonomy metrics is also the 
space of torsion free $G_2$-structures and $H^2_c(X_{ADE}) \cong \mathfrak{h}_{ADE}$ and $H^1(T^3) \cong \mathbb{R}^3$.
The Weyl group action here constitutes a group of non-identity connected diffeomorphisms since it acts non-trivially on the homology. 

Asides from the Ricci flat metric, the $M$-theory moduli space has another contribution coming from the so-called $C$-field and its moduli space is given by:

\begin{proposition} $H_c^3(T^3 \times X_{ADE}, \mathbb{R/Z}) \cong (T^3 \otimes \mathfrak{h}_{ADE})$ which also has an obvious action by $\mathcal{W}_{ADE}$.
\end{proposition}

\begin{definition}
The local moduli space of $M$-theory on $T^3 \times X_{ADE}$, for fixed $T^3$, is 
$T\mathcal{M}^{M}_{ADE} = H_c^3(T^3 \times X_{ADE}, \mathbb{R/Z}+\sqrt{-1}\mathbb{R})/\mathcal{W}_{ADE} 
\cong ((S^1 \times \mathbb{R})^3 \otimes \mathfrak{h}_{ADE})/\mathcal{W}_{ADE}$.
\end{definition}
The $\sqrt{-1}$ is due to supersymmetry, which requires a complex structure on the moduli space.
By exponentiating these Lie algebra valued fields we obtain:
\begin{proposition}
$\mathcal{M}^M_{ADE} \cong \mathcal{M}_{0}(T^{3},G_{ADE}^{\mathbb{C}})$
\end{proposition}

We now consider the moduli space of $M$-theory on $X= \frac{(T^{3}\times X_{ADE})}{K}$. This requires constructing $K$-invariant Ricci flat metrics and $C$-fields on $(T^{3}\times X_{ADE})$ which requires knowing how $K$ acts on the hyperK\"ahler structure $\vec{\omega}$.
Let us denote the $M$-theory moduli space by $(\mathcal{M}^M_{ADE})^K$ since it is naturally given by the $K$-invariant subsets of $\mathcal{M}^M_{ADE}$.

In the previous section we classified the $K$-invariant subspaces of the rhs of proposition 4. Hence, the same must be true of the $M$-theory moduli space $\mathcal{M}^M_{ADE}$. This, however, implies that the appropriate $K$-invariant Ricci flat metrics exist. In \cite{Acharya:2023xlx} we established that this is the case for $G_{ADE}=\op{SU}(N)$ and $G_{ADE}=\op{Spin}(2N)$, but not explicitly for $G_{ADE}=E_{6,7,8}$. For example, for $G_{ADE}=\op{SU}(N)$, the hyperK\"ahler metrics are given by a Gibbons-Hawking ansatz on circle bundles over $\mathbb{R}^3 \backslash \{ p_1, p_2, ...,p_N\}$ where the $N$-points $p_i$ determine where the circle fibers degenerate to zero size. To obtain a $K$-invariant metric, one must consider $K$-invariant configurations of points in $\mathbb{R}^{3}$, where $K$ acts linearly as $K \subset SO(3)$. These are in one-to-one correspondence with homomorphisms from $K$ to $\mathcal{W}_{A_N}\cong S_N$. A similar construction establishes the analogous statement for $G_{ADE}=\op{Spin(2N)}$ \cite{Acharya:2023xlx}. Hence, we have proven that
\begin{theorem}
    $(\mathcal{M}^M_{ADE})^K \cong \mathcal{M}_{0}(T^{3},G_{ADE}^{\mathbb{C}})^K$ for $G_{ADE}=\op{SU}(N)$ and $G_{ADE}=\op{Spin}(2N)$
\end{theorem}
This establishes for these cases that the $M$-theory moduli spaces and the space of flat connections on $M^3_K$ satisfying the conditions of theorem 1 are isomorphic.
We leave it as a conjecture for $G_{ADE}=E_{6,7,8}$.

\bibliographystyle{plain}
\bibliography{References}
\end{document}